\documentclass[
showpacs,
floatfix,
aps,
prb,
amsmath,
twocolumn,
groupaddress,
footinbib,
]{revtex4-1}
\usepackage{amsmath,amssymb}
\usepackage{amscd, latexsym}
\usepackage{mathrsfs}
\usepackage{graphicx}
\usepackage{epstopdf}
\usepackage{cancel}
\usepackage{amsfonts}
\usepackage{exscale}
\usepackage{dcolumn}
\usepackage{bm}
\usepackage{color}
\usepackage{natbib}

\usepackage{lmodern}
\usepackage[T1]{fontenc}
\usepackage[latin9]{inputenc}
\usepackage{verbatim}
\usepackage{float}
\usepackage[unicode=true,pdfusetitle,
 bookmarks=true,bookmarksnumbered=false,bookmarksopen=false,
 breaklinks=false,pdfborder={0 0 0},backref=false,colorlinks=true]
 {hyperref}
\usepackage{breakurl}

\begin{document}

\title{Full Counting Statistics of Photons Emitted by Double Quantum Dot}

\author{Canran Xu}
\author{Maxim G. Vavilov}
\affiliation{Department of Physics, University of Wisconsin-Madison, Wisconsin 53706, USA }

\date{\today}

\begin{abstract}

We analyze the full counting statistics of photons emitted by a double quantum dot (DQD) coupled to 
a high-quality microwave resonator by electric dipole interaction.  We 
show that at the resonant condition between the energy splitting of the DQD and the photon energy in the resonator, photon statistics exhibits both a sub-Poissonian distribution and antibunching. In the ideal case, when the system decoherence stems only from photodetection, the photon noise is reduced below one-half of the noise for the Poisson distribution and is consistent with current noise. The photon distribution remains sub-Poissonian even at moderate decoherence in the DQD. We demonstrate that Josephson junction based photomultipliers can be used to experimentally assess statistics of emitted photons.

\end{abstract}

\pacs{73.23.-b,  73.63.Kv,   42.50.Ar}

\maketitle

\section{Introduction}
The statistics of photons emitted by an electric current depends on the electron state of a conductor.
If the electric current were classical, the photon field would be in a coherent state\cite{Glauber1963}  with  Poissonian statistics.
An electron system with strong inelastic processes is characterized by thermal distribution and produce black-body radiation with super-Poissonian  statistics of emitted photons.
However, if the electron distribution is  far from equilibrium, the photon counting statistics may become 
sub-Poissonian\cite{Beenakker2001,*Beenakker2004, *Lebedev2010}.

Several experiments have recently been developed to study the statistics of photons in the GHz frequency range.  Experiments\cite{Schuster2007a,*Hofheinz2008,Gabelli2004} measured the photon statistics in a steady state of high quality resonator and distinguished between the thermal source and a coherent drive. The photon noise of a quantum point contact at finite bias was also investigated using an amplifier\cite{Zakka-Bajjani2007}.
An alternative approach to study photon statistics utilizes a photon counter\cite{Chen11,*Poudel2012}.
An individual photon counter can provide the statistics of emitted photons, while a system with two counters can be used for Hanbury Brown-Twiss (HBT) interferometry\cite{Carmichael2008}, \textit{e.g.} measurement of the second-order intensity correlation function $g_{\rm}^{(2)}(\tau)$. Generic HBT measurement indicates that noninteracting bosons and fermions would exhibit bunching and antibunching, respectively\cite{Jeltes2007}, while several more complicated examples of photon statistics caused by quantum electron transport have been proposed in systems with quantum point contact\cite{Beenakker2001,*Beenakker2004,  Lebedev2010} and quantum Hall regime\cite{Ikushima2011}.

\begin{figure}
\begin{centering}
\includegraphics[width=1\columnwidth]{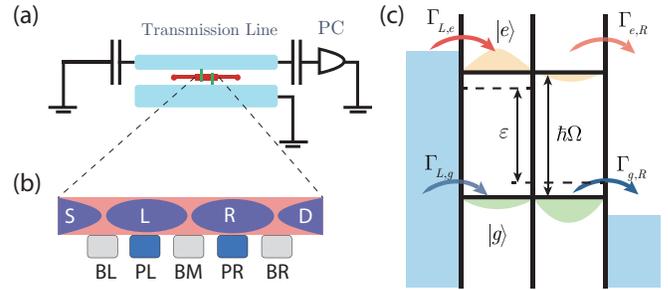}
\end{centering}
\caption{\label{fig:scheme}(Color online) 
(a) An illustration of a DQD and a resonator (a $\lambda/2$--transmission line) coupled to a photon counter (PC). 
(b) In  the DQD, electrons are confined to the left (L) and right (R) dots by barrier gates BL, BM, and BR that also control electron tunneling rates between the source, S, and the left dot, the left and right dots, and the right dot and the drain, D, respectively. 
Electrostatic energies of two quantum dots are defined by the plunger gates, PL and PR,  and the PL gate is also connected to an antinode of the transmutation line. (c) Electronic eigenstates of the DQD and tunneling from the left lead to the ground/excited state with rate $\Gamma_{L,g/e}$, and from the ground/excited state to the right lead with rate $\Gamma_{g/e,R}$.}
\end{figure}

In this paper we study statistics of photon radiation from a DQD coupled to a microwave resonator, a system that was recently studied experimentally by several groups\cite{Delbecq2011,*Frey2012a,*Petersson2012,*Toida2013}. We show that photon statistics is sub-Poissonian with reduced noise in the flux of emitted photons from the resonator. This regime of reduced noise is robust for the considered system.  While it is widely expected that photons produced by an electron source may show sub-Poissonian statistics, such a regime usually occurs under several stringent conditions. 
In particular, the emission statistics of a quantum point contact is sub-Poissonian only if the voltage bias does not exceed twice the photon energy and the contact has a single conduction channel, otherwise photons have super-Poissonian distribution\cite{Beenakker2001,*Beenakker2004,Lebedev2010}. In the setup considered here, a combination of Fermi statistics and repulsion of electrons maintains the reduced noise in photon flux in a wide range of system parameters.  Even a short dephasing time which is the main constraint for observation of quantum effects in DQDs hardly  changes the distribution of emitted photons.  It is the energy relaxation processes in the DQD that drive the photon distribution from sub- to super--Poissonian as the relaxation causes equilibration of the whole system and brings the photon field to a state similar to that of a thermal radiation.  We argue  that the system of a coupled DQD and a resonator can be used to study a cross-over from non-equilibrium to a thermal state in an interacting quantum system.  Finally, our analysis indicates that the Josephson-based photon counters \cite{Chen11,*Poudel2012} are suitable for studies of photon emission statistics by a DQD.

\section{Counting Statistics Formalism}\label{sec:III}

\subsection{Double quantum dot coupled to a resonator}

We study the statistical properties of photon emission by a voltage--biased DQD.
The Hamiltonian for a system of a DQD and a resonator, shown in Fig.~\ref{fig:scheme}, is presented as a combination of three terms,  $H=H_{\text{DQD}}+H_{\text{ph}}+H_{\text{int}}$.  The Hamiltonian of a DQD in the Coulomb blockade regime near a triple point in its electrostatic stability diagram\cite{VanderWiel2002} is represented by
\begin{equation}
H_{\text{DQD}}=\frac{1}{2}\hbar \varepsilon\tau_{z}+\hbar\mathcal{T}\tau_{x},
\label{eq:1}
\end{equation}
in the basis of electron states in the left, $|L\rangle$, and right, $|R\rangle$, quantum dots with electrostatic energy $\varepsilon$ and the tunneling amplitude
${\cal T}$; in this basis, $\tau_{z}=\left|L\right\rangle \left\langle L\right|-\left|R\right\rangle \left\langle R\right|$ and
$\tau_{x}=\left|R\right\rangle \left\langle L\right|+\left|L\right\rangle \left\langle R\right|$.
The term $H_{\text{ph}}=\hbar\omega_0a^{\dagger}a$ represents a noninteracting photon mode in the resonator.
The interaction between charge and photon degrees of freedom is described by the Hamiltonian $H_{\text{int}}=\hbar g_{0}\left(a^{\dagger}+a\right)\tau_z$\cite{Childress2004,Jin2011,Jin2012a,Xu2013}.

In further calculations, we use the eigenstates
of the DQD Hamiltonian, Eq.~\eqref{eq:1}, namely the ground, $|g\rangle$, and excited,
$|e\rangle$, states:
\begin{equation}
\begin{split}
\left|e\right\rangle  & =\cos(\theta/2)\left|L\right\rangle +\sin(\theta/2)\left|R\right\rangle , \label{eq:eigenbasis}   \\ 
\left|g\right\rangle  & =-\sin(\theta/2)\left|L\right\rangle +\cos(\theta/2)\left|R\right\rangle .
\end{split}
\end{equation}
Here $\theta=\arctan(2\cal{T}/\varepsilon)$ characterizes the hybridization
between states $|L\rangle$ and $|R\rangle$ due to inter dot tunneling. The energy splitting
between the eigenstates $\hbar \Omega =\hbar\sqrt{\varepsilon^{2}+4{\cal T}^{2}}$
can be tuned via gate voltages.\cite{VanderWiel2002} In the eigenstate basis, Eq.~\eqref{eq:eigenbasis}, and within rotating wave approximation, the Hamiltonian $H$ is
\begin{equation}
H=\frac{\hbar\Omega}{2}\sigma_{z}+\hbar\omega_0a^{\dagger}a+\hbar g(a^{\dagger}\sigma^{-}+a\sigma^{+}),\label{eq:Hamiltonian}
\end{equation}
where $g=g_{0}\sin\theta$ is the effective electron--photon coupling constant, and  
$\sigma^{-}=\left|g\right\rangle \left\langle e\right|$, $\sigma^{+}=\left|e\right\rangle \left\langle g\right|$. 

We analyze the behavior of the system with Hamiltonian Eq.~\eqref{eq:Hamiltonian}
in the presence of decoherence in electron and photon degrees of freedom  and tunneling of electrons between the DQD and the leads by employing the master equation for the full density matrix 
\begin{equation}
\dot{\rho}=
{\cal L}\rho=
-\frac{i}{\hbar}\left[H,\rho\right]+{\cal D}_{\rm tot}\rho,\label{eq:drho}
\end{equation}
where the commutator describes the
unitary evolution of the system and 
\begin{equation}
{\cal D}_{{\rm tot}}\rho = \kappa{\cal K}(a)\rho+{\cal D}_{\rm DQD}\rho
\end{equation}
accounts for the total
dissipative evolution described by Lindblad superoperators 
\begin{equation}
{\cal K}(x)\rho=\left(2x\rho x^{\dagger}-x^{\dagger}x\rho-\rho x^{\dagger}x\right)/2.
\end{equation} 
The first term, $\kappa{\cal K}(a)\rho$, represents photon detection by an ideal photon counter with rate
$\kappa$. The second term, 
\begin{equation}
{\cal D}_{{\rm DQD}}\rho=\gamma_{r}{\cal K}(\sigma^{-})\rho+\frac{\gamma_{\phi}}{2}{\cal K}(\sigma_{z})\rho+\Gamma_{l}{\cal K}(c_{l}^{\dagger})\rho+\Gamma_{r}{\cal K}(c_{r})\rho,
\end{equation}
describes dissipative dynamics of the DQD. Here,  $\gamma_r{\cal K}(\sigma^{-})\rho$
corresponds to electron relaxation from the excited to ground states at zero temperature; 
$\gamma_{\phi}{\cal K}(\sigma_{z})\rho/2$
represents dephasing with rate $\gamma_{\phi}$;
the last two terms in $\mathcal{D}_{\rm DQD}\rho$ account for the processes\cite{Jin2011}  of loading state $\left|L\right\rangle $ from the source with tunneling rate $\Gamma_l$ 
and unloading state $\left|R\right\rangle $ to the drain with tunneling rate $\Gamma_r$, see Fig.~\ref{fig:scheme}, and we introduced 
 $c_{r}=\left|0\right\rangle \left\langle R\right|$
and $c_{l}^{\dagger}=\left|L\right\rangle \left\langle 0\right|$.

We calculate the full counting statistics (FCS) of emitted photons defined as probability distribution $P_n(t)$ to count $n$ photons during measurement time $t$. For a lossless resonator, the average photon count $\langle n\rangle= \sum nP_n(t)=\kappa \bar N t$  is determined by the photon number $\bar N=\langle a^\dagger a\rangle_{\rm st} = {\rm Tr}\{\rho_{\rm st}a^{\dagger}a\}$ in the resonator  and the photon detection rate\cite{Clerk2010} $\kappa$, 
here $\rho_{\rm st}$ is the steady state solution of Eq.~\eqref{eq:drho}: $\mathcal{L}\rho_{\rm st} = 0$. In particular, we are interested in the Fano factor, $F_{\rm ph}=[\langle n^2\rangle -\langle n\rangle^2]/\langle n\rangle$, that characterizes its noise property. By definition, photon emission is Poissonian if $F_{\rm ph}=1$, while for sub-(super-) Poissonian processes, 
$F_{\rm ph}<1$ ($F_{\rm ph}>1$).

\subsection{Quantum jump approach}

In this section we utilize the quantum jump approach\cite{Gardiner2008} to calculate the FCS. The Liouvillian in Eq.~\eqref{eq:drho} can be decomposed as
\begin{equation}
\dot{\rho}(t)={\cal L}\rho(t)=\left({\cal L}_{0}+{\cal J}\right)\rho(t),
\end{equation}
where we have singled out the jump superopertor, ${\cal J}\rho=\kappa a\rho a^{\dagger}$,
to describe the stochastic quantum jump associated with photon detection,
and ${\cal L}_{0}$ governing the deterministic dynamics of the system.
Since quantum jumps are discretized in counted photon numbers, the full density matrix $\rho(t)$ can be resolved in terms of individual components $\rho^{(n)}(t)$  representing a quantum trajectory with
$n$ photons being counted by the photon detector during time interval $[0, t]$:
\begin{equation}
\rho(t)=\sum_{n}\rho^{(n)}(t).
\end{equation}
By definition, the equation of motions for $\rho^{(n)}(t)$ is
\begin{equation}
\dot{\rho}^{(n)}(t)={\cal L}_{0}\rho^{(n)}(t)+{\cal J}\rho^{(n-1)}(t).\label{eq:supp_3a}
\end{equation}
These equations of motion are coupled
and therefore hard to solve. It is more convenient to define
the generalized density matrix
\begin{equation}
\tilde{\rho}(t,s)=\sum_{n}s^{n}\rho^{(n)}(t),\label{eq:supp_2}
\end{equation}
by introducing the counting variable for photons, $s$. The equations
of motion for $\tilde{\rho}(t,s)$
is obtained by multiplying Eq.~\eqref{eq:supp_3a} by $s^{n}$ and
taking sum over $n$, 
\begin{equation}
\dot{\tilde{\rho}}(t,s)={\cal M}(s)\tilde{\rho}(t,s),\label{eq:supp_3}
\end{equation}
with
\begin{equation}
{\cal M}(s)={\cal L}_{0}+s{\cal J}.\label{eq:supp_5}
\end{equation}
For $s=1$, Eq.~\eqref{eq:supp_3} reduces to the original master equation  
Eq.~\eqref{eq:drho}.
The formal solution of Eq.~\eqref{eq:supp_3} is 
\begin{equation}
\tilde{\rho}(t,s)=e^{{\cal M}(s)t}\tilde{\rho}(0,s),\label{eq:supp_6}
\end{equation}
where the initial state is chosen to be the steady state, $\tilde{\rho}(0,s)=\rho_{\text{st}}$.

Next, we introduce moment generating function 
\begin{equation}
{\cal G}(t,s)=\text{Tr}\left\{\tilde{\rho}(t,s)\right\}=
\text{Tr}\left\{e^{{\cal M}(s)t}\tilde{\rho}(0,s)\right\}.\label{eq:supp_GF}
\end{equation}
This function permits one to calculate the higher order moments. 
Indeed, the  $n$ resolved density matrix allows
us to obtain the FCS of the system by taking trace of $\rho^{(n)}(t)$:
\begin{equation}
P_{n}(t)=\text{Tr}\{\rho^{(n)}(t)\}.\label{eq:supp_7}
\end{equation}
Then, according to Eqs.~\eqref{eq:supp_2} and \eqref{eq:supp_GF}, we identify 
\begin{equation}
{\cal G}(t,s)=\sum_{n}s^{n}P_{n}(t).\label{eq:supp_8}
\end{equation}
The probability distribution $P_{n}(t)$ is given by the inverse Fourier transform in parameter $s=\exp(i\chi)$:
\begin{equation}
P_{n}(t)=\int^{2\pi}e^{-in\chi}{\cal G}(t,e^{i\chi})\frac{d\chi}{2\pi}.
\end{equation}
The factorial moments $\langle\langle n^m \rangle\rangle_{\rm f}$ of $n$ can be obtained by derivatives of ${\cal G}(t,s)$:
\begin{equation}
\langle\langle n^m \rangle\rangle_{\rm f} = \sum_{n}P_n(t)\prod_{i=0}^{m-1}(n-i) = \left.\frac{\partial^{m}{\cal G}(t,s)}{\partial s^{m}}\right|_{s=1}.\label{eq:supp_11}
\end{equation}
We note that Eq.~\eqref{eq:supp_6} is understood as a Dyson series, therefore the generalized density
matrix can be expanded into a sum of  $n$ photon detections:
\begin{align}
\tilde{\rho}(t, s)  =&{\cal S}(t,0)\rho_{{\rm st}}+\sum_{n}\int^{t}dt_{n}\cdots\int^{t_{2}}dt_{1}\label{eq:supp_unravelled} \\ 
 & {\cal S}(t,t_{n})s(t_{n}){\cal J}(t_{n})\cdots{\cal S}(t_{2},t_{1}){\cal J}(t_{1}){\cal S}(t_{1},0)\rho_{{\rm st}},\nonumber 
\end{align}
where ${\cal S}(t_{1,}t_{2})=\exp\left[{\cal L}_{0}(t_{1}-t_{2})\right].$

\subsection{Computation of the Fano factor}\label{sub:Numerical-Method}

In principle, the method described in the previous subsection can be used to calculate $P_n(t)$ and then the Fano factor in terms of the first and second order factorial moments, using Eq.~\eqref{eq:supp_11}.
However, evaluation of factorial
moments involve derivatives of generating function $\mathcal{G}(t,s)$ over
$s$, which is not convenient in practice for numerical
calculations. In this subsection, we describe a numerical method more suitable for numerical evaluation of Fano factors.

As mentioned in the previous subsection,  photon counts over measurement time $t$ are associated with evolution of   the generalized density matrix subject to the corresponding quantum jump $\mathcal{J}$.  Fluctuations in the number of counts are given by 
\begin{equation}
\langle\delta n^2(t)\rangle =\frac{1}{2}\int^{t}_0dt_1 \int^{t}_0dt_2 \left\langle \{\delta \mathcal{J}(t_1),\delta \mathcal{J}(t_2)\}\right\rangle ,
\end{equation}
where $\delta \mathcal{J}(t) = \mathcal{J}(t) - J(t)$ is the quantum fluctuation of the photon counting measurement and $J(t)=\kappa \bar{N}(t)= {\rm Tr}{\cal J}\rho_{\text{st}}$ is the average photon count rate; $\{A,B\}$ stands for an  anticommutator. According to Eq.~\eqref{eq:supp_unravelled}, we take first two orders of derivatives over $s$ and find the correlation function of photon counts during measurement time $t$:
\begin{align}
\langle  n^2(t)\rangle & =\int^{t}dt_1\int^{t}dt_2\left. \frac{\delta^{2} {\rm Tr}\tilde{\rho}(t,s) }{\delta s(t_{1})\delta s(t_{2})}\right|_{s=1}\label{eq:supp_n2_integral}\\ \nonumber 
&+\int^{t}dt\left.\frac{\delta {\rm Tr}\tilde{\rho}(t,s)}{\delta s(t)}\right|_{s=1}\\ \nonumber 
 & =2 \int^{t}dt_1 \int^{t_1}dt_2\left\langle {\cal J}(t_{1}){\cal S}(t_{1},t_{2}){\cal J}(t_{2})
 \right\rangle \\ \nonumber 
 &+  \langle n(t) \rangle 
.\nonumber 
\end{align}
Note that we have implied $t_{1}\geq t_{2}$ in the second line and
thus the term with $t_{1}\leq t_{2}$ is added to symmetrize
the expression with switching on time labels. Then, one can integrate
Eq.~\eqref{eq:supp_n2_integral} with respect to $t_1$ and $t_2$ 
\begin{align}
\langle  \delta n^2\rangle & =
\langle  n^2(t)\rangle-\int^{t}\int^{t}dt_1 dt_2J^{2}
\label{eq:supp_fano} \\ 
 & = \langle n(t) \rangle  +2\int^{t}
d\tau (t-\tau)\left[{\rm Tr}\left({\cal J}\cal S(\tau)\cal J\rho_{{\rm st}}\right)-J^{2}\right]\nonumber \\
 & =\langle n(t) \rangle  +2J^2\int^{t}
d\tau (t-\tau) \left( g_{\rm ph}^{(2)}(\tau)-1\right),\nonumber
\end{align}
using the second order correlation functiont of photon field:
\begin{equation}
g_{{\rm ph}}^{(2)}(\tau)  =\frac{\langle a^{\dagger}a^{\dagger}(\tau)a(\tau)a\rangle}{\langle a^\dagger a \rangle^{2}} 
=\frac{\text{Tr}\left\{ a^{\dagger}ae^{{\cal L}\tau}\left(a\rho_{{\rm st}}a^{\dagger}\right)\right\} }{\text{Tr}\left\{ a^{\dagger}a\rho_{\text{st}}\right\} ^{2}}\label{eq:3}
\end{equation}
Eq.~\eqref{eq:supp_fano} is the famous Mandel's photon counting formula\cite{Zou1990}. Taking into account that $t$ is large compared to the characteristic memory time of the system, Eq.~\eqref{eq:supp_fano} reduces to the expression for the photon Fano factor, independent of $t$  \cite{Zou1990,Kelley1964}
\begin{equation}
F_{\rm ph}=\frac{\langle  \delta n^2\rangle}{\langle n(t) \rangle }=1 +2 J  \int^{\infty}d\tau {\rm Tr}\left\{ g_{\rm ph}^{(2)}(\tau)-1\right\}. \label{eq:F_g2}
\end{equation}

Following Ref. \onlinecite{Flindt2004, Flindt2005}, we introduce the ``Dirac notation'' in Liouvillian
space for steady state $|\text{st}\rangle\rangle\equiv\rho_{\text{st}}$
and a dual vector $\langle\langle e|\equiv\hat{1}$. The inner product
defined in Liouvillian space is the trace over operator ``ket'' in state
``bra''. For example, the inner product of the two former objects
is given by $\langle\langle e|\text{st}\rangle\rangle\equiv\text{Tr}\rho_{\text{st}}=1$.
It is then useful to define the projector ${\cal P}={\cal P}^{2}=|\text{st}\rangle\rangle\langle\langle e|$
onto the steady state as well as its complement ${\cal Q}=1-{\cal P}$.
Note that a useful property of ${\cal P}$ is ${\cal LP}={\cal L}|\text{st}\rangle\rangle\langle\langle e|=0$
and ${\cal PL}=0$, and therefore ${\cal L=QLQ}$.
The propagator ${\cal S}(\tau)$ in the second line of Eq.~\eqref{eq:supp_fano} can be decomposed as ${\cal S}(\tau)={\cal P}+{\cal Q}{\cal S}(\tau){\cal Q}$,
and thus ${\rm Tr}\left({\cal JPJ}\rho_{{\rm st}}\right)=\langle\langle e|{\cal J}|{\rm st}\rangle\rangle\langle\langle e|{\cal J}|{\rm st}\rangle\rangle=J^{2}$.
We obtain
\begin{align}
F_{\rm ph} & =1+\frac{2}{J}\int^{t\rightarrow\infty}d\tau{\cal J}{\cal Q}{\cal S}(\tau){\cal Q}{\cal J}\label{eq:supp_21} \\ 
 & =1-\frac{2}{J}{\rm Tr}\left\{ {\cal J}{\cal QL}^{-1}{\cal Q}{\cal J}\rho_{{\rm st}}\right\}\nonumber \\ 
 & = 1-\frac{2}{J}\langle\langle e|{\cal JRJ}|\text{st}\rangle\rangle,
 \nonumber 
\end{align}
where ${\cal R}={\cal QL}^{-1}{\cal Q}$ is the inverse of the Liouvillian
projected out of the steady state. 

Eq.~\eqref{eq:supp_21} is the key step to evaluate photon Fano factor. We also have to find the inverse of the Loiouvillian, ${\cal R}$,
and project the result out of the steady state. In practice, inverse
of the Liouvillian matrix with large dimension is numerically unstable,
but  we can evaluate the combination $|{\cal W}\rangle\rangle={\cal RJ}|\text{st}\rangle\rangle$
determined by the following equation
\begin{align}
{\cal L}|\mathcal{W}\rangle\rangle&={\cal LRJ}|\text{st}\rangle\rangle={\cal QJ}|\text{st}\rangle\rangle \nonumber \\
&={\cal J}|\text{st}\rangle\rangle-|\text{st}\rangle\rangle\langle\langle e|{\cal J}|\text{st}\rangle\rangle,\label{eq:supp_22}
\end{align}
where the second equality is obtained by the relation ${\cal LR}={\cal LQL}^{-1}{\cal Q}={\cal L}(1-{\cal P}){\cal L}^{-1}{\cal Q}={\cal Q}$. To this end, the solution $|\cal W\rangle\rangle$
in Eq.~\eqref{eq:supp_22} is equivalent to the inverse ${\cal R}$\cite{Flindt2004}. At the end of the calculation
we fix the solution by projection out of the steady state by
condition $\langle\langle e|{\cal W}\rangle\rangle =\text{Tr}\left\{ {\cal R}{\cal J}\rho_{\text{st}}\right\}  =0$, accomplished
by premultiplication of projector ${\cal Q}|{\cal W}\rangle\rangle$.

\begin{figure}
\begin{centering}
\includegraphics[width=1\columnwidth]{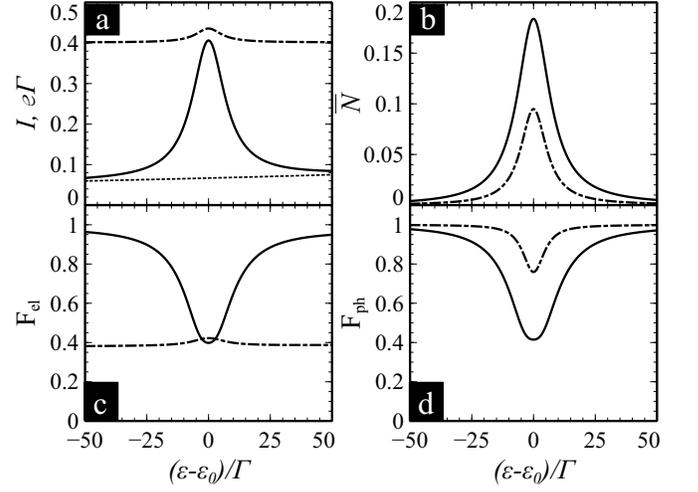}
\end{centering}
\caption{\label{fig:2}(Color online) 
Dependences of (a) the electric current $I$ through the DQD, (b) the average photon number $\bar N$ in the resonator, (c) the electron Fano factor $F_{\rm el}$ and (d) the Fano factor $F_{\rm ph}$ for emitted photons are shown as functions of the electrostatic bias $\hbar\varepsilon$  near the resonance at $\varepsilon_0=-\sqrt{\omega^2_0-4{\cal T}^2}$.
Solid (dashed) lines represent the case for  $\gamma_r=0$ ($\gamma_r=2 \Gamma$). Other system parameters are $\Gamma_{l,r}=\Gamma$, $\omega_0=800\Gamma$, ${\cal T}=200\Gamma$, $g_0=5\Gamma$, $\kappa=2\Gamma$ and $\gamma_{\phi}=0$. Dotted line in panel (a) refers to elastic electric current through a non-interacting DQD.}
\end{figure}

\subsection{Charge full counting statistics}
Charge FCS through a double quantum dot was studied earlier in Refs.~\onlinecite{Blanter2000,*Levitov2003,Jin2012a,Emary2012,Bagrets2003,Flindt2005,Flindt2004,Brandes2008,Marcos2010}. Here  we provide a quick review of the relations for the electric current and current noise through the DQD 
using the master equation formalism. 
The current operator is defined as $\hat{I}=e\Gamma_{r}|R\rangle\langle R|=e\Gamma_{r}c_{r}^{\dagger}c_{r}$. The dc current is given by the expectation value of $\hat I$ with respect to the steady state solution $\rho_{\rm{st}}$ for the density matrix,
\begin{equation}
I=e\Gamma_{r}{\rm Tr} \{\left|R\right\rangle \left\langle R\right|\rho_{\rm st} \}.
\label{eq:current}
\end{equation}

The spectral density of the current fluctuations is defined by the relation
\begin{subequations}\label{eq:S1}
\begin{eqnarray}
&& S(\omega) = \int_{-\infty}^{\infty}\langle\langle 
\hat{I}(t)\hat{I}(t+\tau)\rangle\rangle e^{i\omega t} dt,  \\
&& \langle\langle \hat{I}(t)\hat{I}(t+\tau)\rangle\rangle = \left\langle \hat{I}(t)\hat{I}(t+\tau)\right\rangle -I^2.
\label{eq:I2}
\end{eqnarray}
\end{subequations}
The first term in Eq.~\eqref{eq:I2}
accounts for the concurrence of two electrons at times $t$ and $t+\tau$
\begin{equation}
\langle \hat{I}(t)\hat{I}(t+\tau)\rangle =I^{2}g_{\text{el}}^{(2)}(t,\tau)+eI\delta(\tau)\label{eq:12},
\end{equation}
where the second order correlation function $g_{\text{el}}^{(2)}(\tau)$ is given by 
\begin{equation}
g_{\text{el}}^{(2)}(\tau)=\frac{\text{Tr}\left\{ c_{r}^{\dagger}c_{r}e^{{\cal L}\tau}\left(c_{r}\rho_{\rm st}c_{r}^{\dagger}\right)\right\} }{\text{Tr}\left\{ c_{r}^{\dagger}c_{r}\rho_{\rm st}\right\} ^{2}}.\label{eq:3}
\end{equation}
The last term in Eq.~\eqref{eq:12} represents 
counting the same electron at $t$ and $t+\tau$.

Using Eqs.~\eqref{eq:S1} and \eqref{eq:3}, we can write the current noise spectral function in the form:
\begin{equation}
S(\omega)=I^{2}\int_{-\infty}^{\infty}\left(g_{\text{el}}^{(2)}(\tau)-1\right)e^{i\omega\tau}d\tau+eI.\label{eq:noise_spectrum}
\end{equation}
This result shows that the second order correlation function $g_{\text{el}}^{(2)}(\tau)$ is related to the current noise in both frequency and time
domains. In particular, the Fano factor $F_{\rm el}$ of the charge current that characterizes the low frequency 
limit of $S(\omega)$ is\cite{Emary2012}
\begin{equation}
F=\frac{S(0)}{eI}=1+\frac{2I}{e}\int^{\infty}\left(g_{\text{el}}^{(2)}(\tau)-1\right)d\tau. 
\label{eq:14}
\end{equation}

The formalism discussed above can be further generalized to study cross-correlation functions of electron charge transfer and photon emission\footnote{to be discussed elsewhere.}  Experimental observation of such cross--correlations is a challenging task, but can be achieved by combining charge sensing measurements\cite{Gustavsson2006} and photon detection.\cite{Chen11}

\section{Ideal Photon Counter}
In this section we consider a system consisting of an ideal photon counter and a DQD with equal tunneling rates through left and right contacts, $\Gamma_{l,r}=\Gamma$, and the interdot tunneling amplitude ${\cal T}=200\Gamma$. 
We take $\omega_0=800\Gamma$, and the electron-photon bare coupling $g_0=5\Gamma$. 

\begin{figure}
\begin{centering}
\includegraphics[width=1\columnwidth]{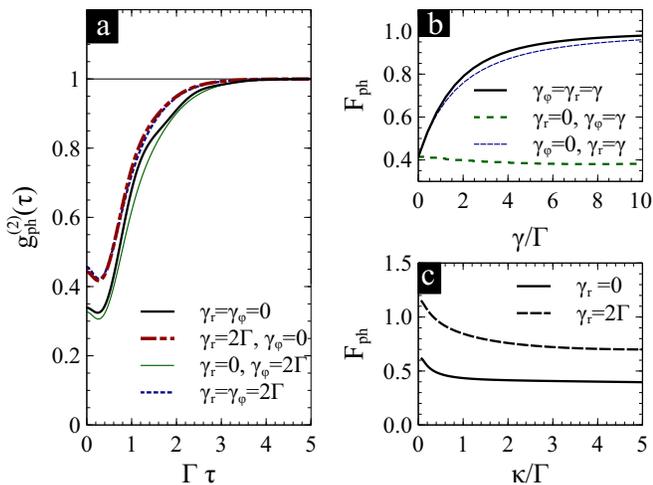}
\end{centering}
\caption{\label{fig:3}(Color online) 
(a) The second order correlation function $g_{\rm ph}^2{(2)}(\tau)$ for photons as a function of time $\tau$ at resonant condition $\Omega=\omega_0$ for different values of energy, $\gamma_r$, and phase, $\gamma_\phi$, relaxation rates. (b) The photon Fano factor $F_{\rm ph}$ as a function of relaxation rate $\gamma$ in three cases $\gamma_r=\gamma_\phi=\gamma$ (solid line), $\gamma_r=\gamma$ and $\gamma_\phi=0$ (dotted line), $\gamma_\phi=\gamma$ and $\gamma_r=0$ (dashed line). (c) $F_{\rm ph}$ as a function of photon detection rate $\kappa$ shows a flat behavior for $\kappa\gtrsim \Gamma$.
Other system parameters in both panels are $\Gamma_{l,r}=\Gamma$, $\omega_0=800\Gamma$, ${\cal T}=200\Gamma$, $g_0=5\Gamma$, $\kappa=2\Gamma$  and  $\gamma_{\phi}=0$[for (a) and (b)].}
\end{figure}

In FIG.~\ref{fig:2} we present (a) dependence of electric current $I=e\Gamma_r\langle c_r^{\dagger}c_r\rangle_{\rm st}$, (b) the average number of photons in the resonator, $\bar N = \langle a^\dagger a\rangle_{\rm st}$, (c) the Fano factor of electronic current, $F_{\rm el}$, and (d) the Fano factor for photon flux, $F_{\rm ph}$. For the former two quantities, we evaluate $\rho_{\rm st}$ and take corresponding expectation values. Solid lines in FIG.~\ref{fig:2} are evaluated for an ideal quantum dot with 
$\gamma_r=\gamma_\phi=0$. In this case, the amplitudes of electric current and the photon flux have a well pronounced peak at the resonant condition $\Omega = \omega_0$, while away from the resonance, photon production is suppressed, $\bar N\to 0$, and the current approaches $I_{0}=e\mathcal{T}^2\Gamma/(\varepsilon^2+3\mathcal{T}^2)$ (a dotted line in FIG.~\ref{fig:2}a) 
for elastic electron transfer through a DQD decoupled from the resonator~\cite{Elattari2002}. Fano factors for both electric current and photon flux are reduced below 1/2, indicating sub-Poissonian statistics with strong suppression of charge and photon noise at the resonance.

Inelastic relaxation facilitates electron transfer through the DQD and increases the electric current above $I_{0}$ even far away from the resonance, $\left|\Omega - \omega_0\right|\gg\Gamma$.  In the presence of such background current,  only a weak enhancement of the current occurs at $\Omega = \omega_0$.  
The electron Fano factor is reduced below unity for $\gamma_r\neq 0$, and a resonant electron transfer with photon emission does not significantly affects $F_{\rm el}$, see FIG~\ref{fig:2}c.  
We observe a resonant emission in the photon flux, see dashed line in FIG.~\ref{fig:2}b, but 
the photon Fano factor, $F_{\rm ph}$, is closer to the value for the Poissonian statistics, $F = 1$.

\begin{figure}
\begin{centering}
\includegraphics[width=1 \columnwidth]{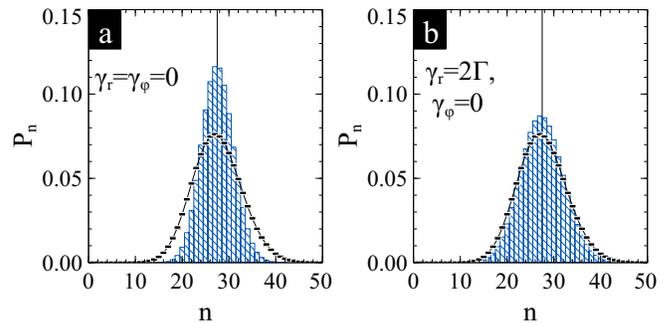}
\end{centering}
\caption{\label{fig:4}(Color online) 
Probabilities $P_n$ to have $n$ emitted photons during time $t$ at resonance $\Omega=\omega_0$ for (a) an ideal DQD without decoherence of electronic states, $\gamma_r=\gamma_\phi=0$ and $t=75/\Gamma$; (b) a DQD with inelastic relaxation $\gamma_r=2\Gamma$, $\gamma_\phi=0$ and $t=145/\Gamma$. Other system parameters in both panels are $\Gamma_{l,r}=\Gamma$, $\omega_0=800\Gamma$, ${\cal T}=200\Gamma$, $g_0=5\Gamma$ and $\kappa=2\Gamma$. A thin curve in both panels represents the corresponding Poisson distribution $P^{(P)}_n=e^{-\bar n}
\bar n^n/n!$ with $\bar n=\sum_n nP_n$ equal to the average number of emitted photons. }
\end{figure}

To evaluate the second order correlation function $g_{\rm ph}^{(2)}(\tau)$ shown in FIG.~\ref{fig:3}a, we compute $\rho_{\rm st}$ and diagonalize the total Liouvillian superoperator $\mathcal{L}$ to obtain $\exp{(\mathcal{L}\tau)}$. The thick solid line shows $g_{\rm ph}^{(2)}(\tau)$ for an ideal DQD, 
$\gamma_{\rm r}=\gamma_\phi=0$.  The probability to observe two photons simultaneously is reduced, $g_{\rm ph}^{(2)}(0)<1$, indicating photon antibunching.
As $\tau$ becomes longer than $\sim 1/\Gamma$, function  $g_{\rm ph}^{(2)}(\tau)$ increases  and eventually approaches its asymptote, $g^{(2)}_\text{ph}(\tau\to\infty)=1$. The integral in Eq.~\eqref{eq:F_g2} with such $g_{\rm ph}^{(2)}(\tau)$ is negative and $F_{\rm ph}<1$  (sub-Poissonian)~\cite{Note1}. 

In the presence of inelastic relaxation in the DQD, $g_{\rm ph}^{(2)}(0)$ increases and  $g_{\rm ph}^{(2)}(\tau)$ reaches its long-time asymptotic value 1
at a shorter time scale.   
However, pure dephasing, $\gamma_\phi$, does not significantly change the shape of $g_{\rm ph}^{(2)}(\tau)$, because individual photon emission is phase-destructive. We also investigate the dependence of $F_{\rm ph}$ on inelastic, $\gamma_r$, and dephasing, $\gamma_{\phi}$ in FIG.~\ref{fig:3}b.  
The inelastic relaxation, $\gamma_r=\gamma$ and $\gamma_\phi=0$, recovers $F_{\rm ph}$ to its Poissonian value, $F_{\rm ph}=1$, as a consequence of the reduced memory of the system, $g^{(2)}_\text{ph}(\tau)\to1$, due to inelastic relaxation.
Pure dephasing, $\gamma_\phi=\gamma$ and $\gamma_r=0$, has weak effect on $F_{\rm ph}$ even for large values of $\gamma_\phi$, as dephasing does not change $g_{\rm ph}^{(2)}(\tau)$ to modify the integral in Eq.~\eqref{eq:F_g2} for $F_{\rm ph}$. The addition of dephasing to relaxation, $\gamma_\phi=\gamma_r=\gamma$, makes no significant corrections to  $F_{\rm ph}$ when compared to $F_{\rm ph}(\gamma_r, 0)$.

\begin{figure}
\includegraphics[width=1 \columnwidth]{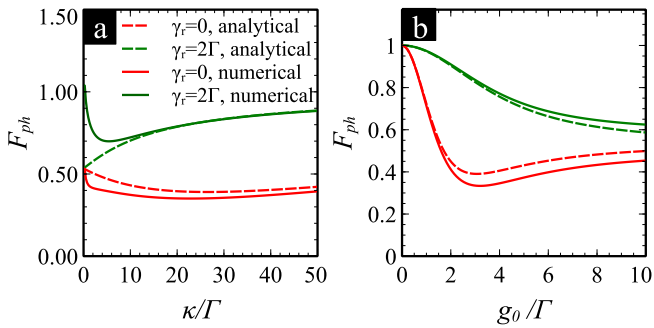}
\caption{\label{fig:(color-online)-Comparison}(Color online) Dependence of
photon noise Fano factor on (a) photon detection rate $\kappa$ and (b) the bare coupling constant $g_0$. In panel (a), the bare coupling constant is fixed $g_{0}=5\Gamma$ and in panel (b) the photon detection rate is fixed $\kappa = 10 \Gamma$. Solid (dashed) lines represent results
for numerical (analytical) calculations with method described in Sec.~\ref{sub:Numerical-Method}(Sec.~\ref{sec:IV}). Other system parameters are
$\Gamma_{l,r}=\Gamma$, $\omega_0=800\Gamma$, ${\cal T}=200\Gamma$.
Both plots indicate that the analytical results agree with numerical
in the limit $\kappa\gtrsim g_{0}, \Gamma$.}
\end{figure}

The dependence of the photon Fano factor on the photon 
detection rate $\kappa$ is also studied. As  $\kappa$ decreases, the average photon number in the resonator increases. 
At large $\bar N$,  photons already present in the resonator cause stimulated emission by the DQD\cite{Astafiev2007,Jin2011}. For an ideal DQD without energy relaxation, $\gamma_r=0$, the photon Fano factor grows fast for $\kappa\lesssim\Gamma$; see FIG.~\ref{fig:3}c.  On the other hand, if the energy relaxation in the DQD is significant, the photon Fano factor can exceed unity and the photon field exhibits properties of a thermal state.  In this respect, inelastic processes in the DQD  enhance photon noise. For a strong photon detection rate, $\kappa\gtrsim \gamma_r, g$, and consequently a low photon number in the resonator, the back-action of the photon field on electrons is negligible. 

In Sec \ref{sec:IV} we apply the adiabatic elimination method\cite{Carmichael2008, Sanchez2007, *Sanchez2008} to study photon statistics emitted by a quantum dot.
 
Next, we present the distribution function of the photon counts $n$ over time $t$.  For this purpose, we compute $\tilde{\rho}(t,s)$ with matrix exponent $\exp{[\mathcal{M}(s)t]}$ and integrate over counting field $\chi$, see Eq.~\eqref{eq:supp_GF} and Eq.~\eqref{eq:supp_11}. For $\gamma_{\rm r}=\gamma_{\phi}=0$, we take $t=75/\Gamma$ and  the distribution $P_n(t)$ is shown in FIG.~\ref{fig:4}a.  
The average value of photon counts, $\langle n\rangle = \sum_n nP_n \simeq 27.6$  is consistent with $\langle n\rangle =\bar N \kappa t$ with $\bar N=0.184$ and their
variance $\langle n^2 \rangle - \langle n \rangle^{2} \simeq  11.6$, in turn gives $F_{\rm ph} = 0.42$, both of which coincide with previous calculation, see Fig.~\ref{fig:2}b,d at $\varepsilon=\varepsilon_0$. 
We present the Poisson distribution with the same expectation value $\langle n \rangle$ by narrow black dots in  FIG.~\ref{fig:4}a. 
$P_n$ is closer to the Poisson distribution for a system with relaxation $\gamma_{\rm r}=2\Gamma$ [see FIG.~\ref{fig:4}b], where $t=145/\Gamma$, $\langle n\rangle \simeq 27.6$ ($\bar N=  0.095$) and $F_{\rm ph} = 0.76$.

\section{Adiabatic elimination of photon degrees of freedom}\label{sec:IV}

In the limit of strong photon detection rate, $\kappa\gg\gamma,g,\Gamma$, the photon field decays so fast that the density matrix can be approximately factorized as 
\begin{equation}
\rho(t)\simeq\rho_{\text{D}}(t)\left(|0\rangle\langle0|\right),
\end{equation} where $|0\rangle$ is the vacuum state of the photon in the transmission line and $\rho_{\text{D}}(t)$ is the reduced density matrix of the DQD.
Thus we can adiabatically eliminate the photon mode, and obtain the
equation of motion for the reduced density matrix $\tilde{\rho}_{\text{D}}(t,s)$
in the interaction picture\cite{Carmichael2008},

\begin{align}
\dot{\tilde{\rho}}_{\text{D}}(t,s)=&\Gamma_{l}{\cal D}(c_{l}^{\dagger})\tilde{\rho}_{\text{D}}(t,s)+\Gamma_{r}{\cal D}(c_{r})\tilde{\rho}_{\text{D}}(t,s)\\ \label{eq:supp_24}
&+\gamma_{*}{\cal D}(\sigma^{-})\tilde{\rho}_{\text{D}}(t,s)+(s-1){\cal J}(\sigma^{-})\tilde{\rho}_{\text{D}}(t,s), \nonumber
\end{align}
where $\gamma_{\text{ph}}=4g^{2}/\kappa$ is the photon-induced
relaxation rate associated with the spontaneous
emission in this large $\kappa$ limit and $\gamma_{*} = \gamma_r + \gamma_{\rm ph}$. Therefore photon absorptions
can be reflected by the jump superoperator ${\cal J}(\sigma^{-})\tilde{\rho}_{\text{D}}(t,s)=\gamma_{\text{ph}}\sigma^{-}\tilde{\rho}_{\text{D}}(t,s)\sigma^{+}$.
In the basis $\rho_{\text{D}}=(\rho_{0},\rho_{g},\rho_{ge},\rho_{eg},\rho_{e})^{T},$
 the matrix ${\cal M}(s)$ in Eq.~\eqref{eq:supp_3} is given by:
\begin{widetext}
\begin{align}
{\cal M}(s) & =\frac{1}{4}\left(\begin{array}{ccccc}
-4\Gamma_{l} & -2\Gamma_{r}\cos\theta+2\Gamma_{r} & -\Gamma_{r}\sin\theta & -\Gamma_{r}\sin\theta & 2\Gamma_{r}\cos\theta+2\Gamma_{r}\\
2\Gamma_{l}\cos\theta+2\Gamma_{l} & 2\Gamma_{r}\cos\theta-2\Gamma_{r} & -\Gamma_{r}\sin\theta & -\Gamma_{r}\sin\theta & 4\gamma_{r}+4s\gamma_{\text{ph}}\\
-2\Gamma_{l}\sin\theta & -\Gamma_{r}\sin\theta & -2\Gamma_{r}-2\gamma_{*} & 0 & -\Gamma_{r}\sin\theta\\
-2\Gamma_{l}\sin\theta & -\Gamma_{r}\sin\theta & 0 & -2\Gamma_{r}-2\gamma_{*} & -\Gamma_{r}\sin\theta\\
-2\Gamma_{l}\cos\theta+2\Gamma_{l} & 0 & -\Gamma_{r}\sin\theta & -\Gamma_{r}\sin\theta & -2\Gamma_{r}\cos\theta-2\Gamma_{r}-4\gamma_{*}
\end{array}\right).
\end{align}
\end{widetext}

To calculate the generating function ${\cal G}(s,t)$ and its derivatives, we take the Laplace
transform of the generalized density matrix, Eq.~\eqref{eq:supp_6},
\begin{equation}
\tilde{\rho}(z,s)_{\rm D}=(z-{\cal M}(s))^{-1}\tilde{\rho}(0,s)_{\rm D}.\label{eq:supp_12}
\end{equation}
Since the long time behavior of the solution is determined by the
residue of the generating function at the pole near $z=0$, i.e.,
${\cal G}(t,s)\sim g(s)e^{z_{0}t}$ with $g(1)=1$, we can expand
the pole around $s=1$: 
\begin{equation}
z_{0}=\sum_{i>0}c_{i}(s-1)^{i},
\end{equation}
and obtain, from Eq.~\eqref{eq:supp_11}, the first two moments,
$\langle\langle n^{i}\rangle\rangle=\langle\left(n-\langle n\rangle\right)^{i}\rangle$:
\begin{align}
\langle\langle n\rangle\rangle & =\left.\frac{\partial g}{\partial s}\right|_{s=1}+c_{1}t,\\
\langle\langle n^{2}\rangle\rangle & =\left.\frac{\partial^{2}g}{\partial s^{2}}\right|_{s=1}-\left[\left.\left(\frac{\partial g}{\partial s}\right)^{2}-\frac{\partial g}{\partial s}\right]\right|_{s=1}+\left(c_{1}+2c_{2}\right)t,
\end{align}
which give the mean and variance of the probability distribution,
respectively. In the asymptotic limit, $t\rightarrow\infty$, all
the information about the moments is included in the expansion coefficients
$c_{i}$. For instance, the Fano factor is given by\cite{Sanchez2007,*Sanchez2008}
\begin{equation}
F\equiv\frac{\langle n^{2}\rangle-\langle n \rangle^2}{\langle n \rangle}=1+\frac{2c_{2}}{c_{1}}.
\end{equation}
To find the coefficients $c_{1}$ and $c_{2}$,
we consider the equation
\begin{equation}
\det\left(z_{0}\hat{1}-{\cal M}(s)\right)=0,\label{eq:supp_26}
\end{equation}
with $z_{0}=c_{1}(s-1)+c_{2}(s-1)^{2}+{\cal O}(s-1)$.
Then we can expand Eq.~\eqref{eq:supp_26} in powers of $s$ and let
the coefficients for each power of $s - 1$ be zero. This procedure generates
a set of equations with $c_{i}$ to arbitrarily large $i$. We provide
two examples below.

First, we consider $\Gamma_{l}=\Gamma_{r}=\Gamma$ and $\theta\rightarrow\pi$
with fixed coupling constant $g$,$ $ the case in which the two levels
in the DQD are weakly overlapping, and we obtain 
\begin{align}
c_{1} & =\frac{\gamma_{\text{ph}}\Gamma}{2\gamma_{*}+\Gamma},\\
c_{2} & =-\frac{\gamma_{\text{ph}}^{2}\Gamma(\gamma_{*}+2\Gamma)}{\left(2\gamma_{*}+\Gamma\right)^{3}}.
\end{align}
The Fano factor is given by 
\begin{equation}
F=1-\frac{2\gamma_{\text{ph}}(\gamma_{*}+2\Gamma)}{\left(2\gamma_{*}+\Gamma\right)^{2}}<1,
\end{equation}
corresponding to the sub-Poissonian noise. When the DQD is tuned to its charge degeneracy, $\theta = \pi/2$, the solutions
then read
\begin{align}
c_{1} & =\frac{\gamma_{\text{ph}}\Gamma(\gamma_{*}+2\Gamma)}{6\gamma_{*}^{2}+11\gamma_{*}\Gamma+4\Gamma^{2}},\\
c_{2} & =-\frac{\gamma_{\text{ph}}^{2}\Gamma(\gamma_{*}+2\Gamma)\left(4\gamma_{*}^{3}+14\gamma_{*}^{2}\Gamma+31\gamma_{*}\Gamma^{2}+20\Gamma^{3}\right)}{\left(6\gamma_{*}^{2}+11\gamma_{*}\Gamma+4\Gamma^{2}\right)^{3}}.
\end{align}
The Fano factor in this case is
\begin{equation}
F=1-\frac{2\gamma_{\text{ph}}\left(4\gamma_{*}^{3}+14\gamma_{*}^{2}\Gamma+31\gamma_{*}\Gamma^{2}+20\Gamma^{3}\right)}{\left(6\gamma_{*}^{2}+11\gamma_{*}\Gamma+4\Gamma^{2}\right)^{2}},
\end{equation}
again, giving the sub-Poissonian noise. In both cases, Fano factors are below 1 for $\gamma_{\rm ph}\neq 0 $, indicating that it is the interaction between photons and electrons that gives rise to the sub-Poissonian statistics.

As mentioned above, this analytical method is valid in the limit when $\kappa$
is large. We hereby make a comparison between analytical and numerical
results, see FIG. \ref{fig:(color-online)-Comparison}. In both plots,
we do not consider dephasing effects. The calculations indicate that the analytical
method presented above agrees with numerical results in the limit
$\kappa\gtrsim g_{0}, \Gamma$.

\section{Photon Counting Statistics Measured By Josephson Photomultipliers}
In this section, we demonstrate that emitted photon statistics can be measured and justified by actual devices. A possible measurement  is based on recently developed 
Josephson photomultipliers (JPM).\cite{Chen11, Govia2012}
In these devices the Josephson coupling $E_J$ dominates over charging energy $E_c$, therefore it is convenient to write the Hamiltonian of the 
JPM in terms of the phase operator $\phi$ across the Josephson junction:
\begin{equation}
H_{\rm JPM} = - \frac{E_c}{2}\frac{d^2}{d\phi^2} - E_J\left(\cos\phi-\frac{I}{I_0}\phi\right),\label{eq:PCJJ}
\end{equation}
where $I$ is the biased current and $I_0$ is the critical current of the junction. For $I\lesssim I_0$, the potential energy takes a ``washbroad''shape, with a few discrete energy levels in the minima separated from the continuum. In this manner, we can tune the biased current such that only two phase states are bounded in the local minima. 

Next we couple the microwave resonator to the JPM with Jaynes-Cummings type interaction. The total Hamiltonian is then written as
\begin{equation}
\tilde{H}=H+\hbar g_{\rm JPM}(a^{\dagger}\upsilon^{-}+a\upsilon^{+})+\frac{\hbar\omega_{\rm JPM}}{2}\upsilon_{z},
\end{equation}
where $H$ is the Hamiltonian Eq.~\eqref{eq:Hamiltonian}, $g_{\rm JPM}$  the coupling constant
between the JPM and a resonator mode, and the Pauli matrices ${\upsilon}^\pm$ and ${\upsilon}_z$
are defined in the basis of eigenstates of the JPM spanned by $|E\rangle$
and $|G\rangle$. In a similar manner, the dynamics of the system
is governed by the following master equation
\begin{subequations}
\begin{eqnarray}
\dot{\rho} & = & {\cal L}{\rho}=-\frac{i}{\hbar}[\tilde{H},{\rho}]+{\cal D}_{{\rm tot}}{\rho},
\\
{\cal D}_{{\rm tot}}{\rho} & = & \kappa_{0}{\cal K}(a){\rho}+{\cal D}_{{\rm {\rm DQD}}}{\rho}+{\cal D}_{{\rm {\rm JPM}}}{\rho},
\end{eqnarray}
where 
\begin{equation}
{\cal D}_{{\rm JPM}}{\rho}=\gamma_{\rm t}{\cal K}\left(|V\rangle\langle E|\right){\rho}+
\gamma_{{\rm d}}{\cal K}(\upsilon^{-}){\rho}+
\gamma_{{\rm cap}}{\cal K}(|G\rangle\langle V|){\rho}
\end{equation}
\end{subequations}
with $|V\rangle$ referring to a voltage regime of the junction, $\gamma_{\rm t}$ is the tunneling to this regime from the excited state and $\gamma_{{\rm cap}}$ is the capturing rate from the voltage regime to the ground state.  By defining state $|V\rangle$, we imply that the junction is out of the Hilbert space of two localized states $|G\rangle, |E\rangle$ near a local minima of the Josephson energy and does not correspond to a particular quantum state.  While the junction evolves out of these two states, a finite voltage develops  that can be identified by the measurement circuit as a photon detection signal.  Then, the dissipation in the circuit leads to recapturing the junction in its ground state and the operation cycle closes. 
%
%
%
%
%
For this measurement scheme, we use the similar formalism introduced in Section~\ref{sec:III} to calculate the Fano factor and FCS associated with the jump operator $\mathcal{J} {\rho} = \gamma_{\rm t} |V\rangle\langle E| {\rho} |E\rangle\langle V|$.

\begin{figure}
\begin{centering}
\includegraphics[width=0.85\columnwidth]{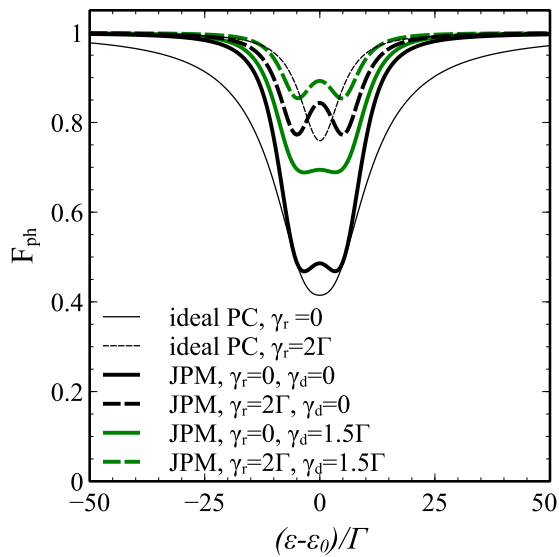}
\end{centering}
\caption{\label{fig:6}
(Color online) Comparison of Fano factor measured by an ideal photon detector and JPM with the same set of parameters of a DQD. Other additional parameters for JPM model are $g_{\rm JPM}= 5\Gamma$, $\omega_{\rm JPM} = \omega_0$, $\kappa_0 = 0.6\Gamma$,  $\gamma_{\rm cap} = 1.5\Gamma$ and $\gamma_{\rm t} = 1.5\Gamma$.  When we introduce the energy relaxation rate $\gamma_d=1.5\Gamma$ 
for the JPM, the resultant Fano factor (green lines) increase towards unity.}
\end{figure} 

\begin{figure}
\begin{centering}
\includegraphics[width=1\columnwidth]{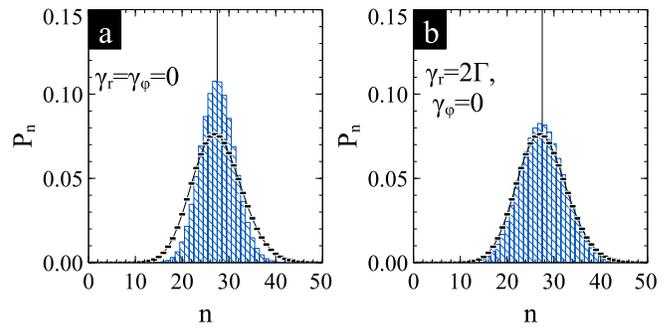}
\end{centering}
\caption{\label{fig:7}
(Color online) Probabilities $P_n$ to have $n$ JJ recordings during time $t$ at resonance $\Omega=\omega_0$ for (a) an ideal DQD without decoherence of electronic states, $\gamma_r=\gamma_\phi=0$ and $t=132/\Gamma$; (b) a DQD with inelastic relaxation $\gamma_r=2\Gamma$, $\gamma_\phi=0$ and $t=293.3/\Gamma$. Other system parameters in both panels are $\Gamma_{l,r}=\Gamma$, $\omega_0=800\Gamma$, ${\cal T}=200\Gamma$, $g_0=5\Gamma$ and $\kappa=2\Gamma$. A thin curve in both panels represents the corresponding Poisson distribution $P^{(P)}_n=e^{-\bar n}
\bar n^n/n!$ with $\bar n=\sum_n nP_n$ equal to the average number of emitted photons. }
\end{figure}

Black thick lines in FIG.~\ref{fig:6} correspond to
the JPM measurement without energy relaxation and thin line are taken from FIG.~\ref{fig:2}d
and correspond to the Fano factor measured by an ideal photon counter for the same set of parameters of the DQD. 
Even though Rabi splitting appears due to coupling between JPM and microwave mode, a fairly good agreement between the ideal  and JPM models of the photon counter indicates that sub-Poissonian statistics of photon emission by a DQD can be experimentally observed. The green lines in FIG.~\ref{fig:6} represent the measured Fano factor with energy relaxation rate $\gamma_d = 1.5 \Gamma$ in the JPM, indicating that energy relaxation of the measurement device would spoil the noise characteristics. This is reminiscent of reduction of quantum efficiency of measurement device\cite{Poudel2012}. The FCS of JPM recordings is shown in Fig.~\ref{fig:7} for cases $\gamma_r = {0, 2\Gamma}$, both of which agree with Fano factor calculation in Fig.~\ref{fig:6}.

\section{Conclusions}
We investigated the statistics of photons emitted by a biased DQD coupled to a lossless resonator.
We calculated the time correlation function $g^{(2)}_{\rm ph}(\tau)$ and found that photons exhibit antibunching, $g^{(2)}(\tau) < g^{(2)}(\tau \to \infty) = 1$. We also calculated photon counting statistics $P_n(t)$ of observing $n$ photons during a fixed time interval $t$. We find  that distribution $P_n(t)$ shows a sub-Poissonian statistics if measured by an ideal photon counter.  We also demonstrate that photon full counting statistics can be accurately studied experimentally by utilizing a Josephson photomultiplier\cite{Chen11, Govia2012}.

In recent experiments, decoherence rates were comparable to the strength of the electron--photon coupling. For this reason, we investigated the effect on charge and photon statistics of pure dephasing in the DQD and energy relaxation.  We found that pure dephasing does not significantly modify the charge transfer or photon emission statistics, but  the inelastic relaxation processes result in several drastic changes (see FIG.~\ref{fig:2}): (\textit{i}) The electric current and its noise acquire strong background as the inelastic processes facilitate the charge transfer throughout the DQD, and the peak in $I$ and the pit in $F_{\rm el}$ at the resonant condition $E_e-E_g=\hbar \omega_0$ are flattened. (\textit{ii})~The photon number $\bar N$  at the resonance is suppressed as the effective photon source is reduced due to additional channels for the $e\to g$ transition via inelastic events. (\textit{iii})~Photon Fano factor $F_{\rm ph}$ as a function of the level spacing is flatten as well. 
In the presence of inelastic electron relaxation, the memory time of the DQD is reduced, which increases the photon correlation function $g^{(2)}_{\rm ph}$ at short time scales[see Fig.~\ref{fig:3}a] and brings the photon Fano factor to its Poisson value, $F_{\rm ph}=1$, as shown in FIG.~\ref{fig:2}.

\section{Acknowledgements.}
We are grateful to R. McDermott, J. Petta and H. T\"ureci for fruitful discussions.  The work was supported by NSF Grant No. DMR-1105178, ARO and LPS Grant No. W911NF-11-1-0030.

\end{document}